\documentclass[10pt,a4paper,onecolumn]{article}
\usepackage[numbers,sort&compress]{natbib}
\usepackage{marginnote}
\usepackage{graphicx}
\usepackage{xcolor}
\usepackage{authblk,etoolbox}
\usepackage{titlesec}
\usepackage{calc}
\usepackage{tikz}
\usepackage{hyperref}
\hypersetup{colorlinks,breaklinks=true,
            urlcolor=[rgb]{0.0, 0.5, 1.0},
            linkcolor=[rgb]{0.0, 0.5, 1.0}}
\usepackage{caption}
\usepackage{tcolorbox}
\usepackage{amssymb,amsmath}
\usepackage{ifxetex,ifluatex}
\usepackage{seqsplit}
\usepackage{xstring}
\usepackage{doi}
\usepackage{orcidlink}

\usepackage{academicons}
\definecolor{orcidlogocol}{HTML}{A6CE39}

\usepackage[T1]{fontenc}
\usepackage{lmodern}

\usepackage{float}
\let\origfigure\figure
\let\endorigfigure\endfigure
\renewenvironment{figure}[1][2] {
    \expandafter\origfigure\expandafter[H]
} {
    \endorigfigure
}

\usepackage{fixltx2e} 

\let\textttOrig=\texttt
\def\texttt#1{\expandafter\textttOrig{\seqsplit{#1}}}
\renewcommand{\seqinsert}{\ifmmode
  \allowbreak
  \else\penalty6000\hspace{0pt plus 0.02em}\fi}

\makeatletter
\let\href@Orig=\href
\def\href@Urllike#1#2{\href@Orig{#1}{\begingroup
    \def\Url@String{#2}\Url@FormatString
    \endgroup}}
\def\href@Notdoi#1#2{\def\tempa{#1}\def\tempb{#2}%
  \ifx\tempa\tempb\relax\href@Urllike{#1}{#2}\else
  \href@Orig{#1}{#2}\fi}
\def\href#1#2{%
  \IfBeginWith{#1}{https://doi.org}%
  {\href@Urllike{#1}{#2}}{\href@Notdoi{#1}{#2}}}
\makeatother

\newlength{\cslhangindent}
\newlength{\csllabelwidth}
 {
  \setlength{\parindent}{0pt}
  \ifodd #1 \everypar{\setlength{\hangindent}{\cslhangindent}}\ignorespaces\fi
  \ifnum #2 > 0
  \setlength{\parskip}{#2\baselineskip}
  \fi
 }%
 {}
\usepackage{calc}

\usepackage[top=3.5cm, bottom=3cm, right=1.5cm, left=1.0cm,
            headheight=2.2cm, reversemp, includemp, marginparwidth=4.5cm]{geometry}

\titleformat{\section}
  {\normalfont\sffamily\Large\bfseries}
  {}{0pt}{}
\titleformat{\subsection}
  {\normalfont\sffamily\large\bfseries}
  {}{0pt}{}
\titleformat{\subsubsection}
  {\normalfont\sffamily\bfseries}
  {}{0pt}{}
\titleformat*{\paragraph}
  {\sffamily\normalsize}

\usepackage{fancyhdr}
\makeatletter
\let\ps@plain\ps@fancy
\fancyheadoffset[L]{4.5cm}
\fancyfootoffset[L]{4.5cm}

\definecolor{linky}{rgb}{0.0, 0.5, 1.0}

\newtcolorbox{repobox}
   {colback=red, colframe=red!75!black,
     boxrule=0.5pt, arc=2pt, left=6pt, right=6pt, top=3pt, bottom=3pt}

\newcommand{\ExternalLink}{%
   \tikz[x=1.2ex, y=1.2ex, baseline=-0.05ex]{%
       \begin{scope}[x=1ex, y=1ex]
           \clip (-0.1,-0.1)
               --++ (-0, 1.2)
               --++ (0.6, 0)
               --++ (0, -0.6)
               --++ (0.6, 0)
               --++ (0, -1);
           \path[draw,
               line width = 0.5,
               rounded corners=0.5]
               (0,0) rectangle (1,1);
       \end{scope}
       \path[draw, line width = 0.5] (0.5, 0.5)
           -- (1, 1);
       \path[draw, line width = 0.5] (0.6, 1)
           -- (1, 1) -- (1, 0.6);
       }
   }

\patchcmd{\@maketitle}{center}{flushleft}{}{}
\patchcmd{\@maketitle}{center}{flushleft}{}{}
\patchcmd{\@maketitle}{\LARGE}{\LARGE\sffamily}{}{}
\def\maketitle{{%
  
  \AB@maketitle}}
\makeatletter
\renewcommand\AB@affilsepx{ \protect\Affilfont}
\renewcommand\AB@affilnote[1]{{\bfseries #1}\hspace{3pt}}
\renewcommand{\affil}[2][]%
   {\newaffiltrue\let\AB@blk@and\AB@pand
      \if\relax#1\relax\def\AB@note{\AB@thenote}\else\def\AB@note{#1}%
        \setcounter{Maxaffil}{0}\fi
        \begingroup
        \let\href=\href@Orig
        \let\texttt=\textttOrig
        \let\protect\@unexpandable@protect
        \def\thanks{\protect\thanks}\def\footnote{\protect\footnote}%
        \@temptokena=\expandafter{\AB@authors}%
        {\def\\{\protect\\\protect\Affilfont}\xdef\AB@temp{#2}}%
         \xdef\AB@authors{\the\@temptokena\AB@las\AB@au@str
         \protect\\[\affilsep]\protect\Affilfont\AB@temp}%
         \gdef\AB@las{}\gdef\AB@au@str{}%
        {\def\\{, \ignorespaces}\xdef\AB@temp{#2}}%
        \@temptokena=\expandafter{\AB@affillist}%
        \xdef\AB@affillist{\the\@temptokena \AB@affilsep
          \AB@affilnote{\AB@note}\protect\Affilfont\AB@temp}%
      \endgroup
       \let\AB@affilsep\AB@affilsepx
}
\makeatother

\renewcommand\Affilfont{\sffamily\small\mdseries}
\ifnum 0\ifxetex 1\fi\ifluatex 1\fi=0 
  \usepackage[T1]{fontenc}
  \usepackage[utf8]{inputenc}

\else 
  \ifxetex
    \usepackage{mathspec}
    \usepackage{fontspec}

  \else
    \usepackage{fontspec}
  \fi
  \defaultfontfeatures{Ligatures=TeX,Scale=MatchLowercase}

\fi
\IfFileExists{upquote.sty}{\usepackage{upquote}}{}
\IfFileExists{microtype.sty}{%
\usepackage{microtype}
\UseMicrotypeSet[protrusion]{basicmath} 
}{}

\usepackage{hyperref}
\hypersetup{unicode=true,
            pdftitle={HofstadterTools: A Python package for analyzing the Hofstadter model},
            pdfborder={0 0 0},
            breaklinks=true}
\let\addcontentslineOrig=\addcontentsline
\def\addcontentsline#1#2#3{\bgroup
  \let\texttt=\textttOrig\addcontentslineOrig{#1}{#2}{#3}\egroup}
\let\markbothOrig\markboth
\def\markboth#1#2{\bgroup
  \let\texttt=\textttOrig\markbothOrig{#1}{#2}\egroup}
\let\markrightOrig\markright
\def\markright#1{\bgroup
  \let\texttt=\textttOrig\markrightOrig{#1}\egroup}

\usepackage{graphicx,grffile}
\makeatletter
\def\maxwidth{\ifdim\Gin@nat@width>\linewidth\linewidth\else\Gin@nat@width\fi}
\def\maxheight{\ifdim\Gin@nat@height>\textheight\textheight\else\Gin@nat@height\fi}
\makeatother
\setkeys{Gin}{width=\maxwidth,height=\maxheight,keepaspectratio}
\IfFileExists{parskip.sty}{%
\usepackage{parskip}
}{
\setlength{\parindent}{0pt}
\setlength{\parskip}{6pt plus 2pt minus 1pt}
}
\ifx\paragraph\undefined\else
\let\oldparagraph\paragraph
\renewcommand{\paragraph}[1]{\oldparagraph{#1}\mbox{}}
\fi
\ifx\subparagraph\undefined\else
\let\oldsubparagraph\subparagraph
\renewcommand{\subparagraph}[1]{\oldsubparagraph{#1}\mbox{}}
\fi

\title{HofstadterTools: A Python package for analyzing the Hofstadter model}

        \author[1]{Bartholomew Andrews\,\orcidlink{0000-0002-9079-7433}\,}
    
      \affil[1]{Department of Physics, University of California, Berkeley, USA}
  \date{\vspace{-7ex}}

\begin{document}
\maketitle

\marginpar{

  \begin{flushleft}
  \sffamily\small

  {\bfseries DOI:} \href{https://doi.org/10.21105/joss.06356}{\color{linky}{10.21105/joss.06356}}

  \vspace{2mm}

  {\bfseries Software}
  \begin{itemize}
    \setlength\itemsep{0em}
    \item \href{https://github.com/openjournals/joss-reviews/issues/6356}{\color{linky}{Review}} \ExternalLink
    \item \href{https://github.com/HofstadterTools/HofstadterTools}{\color{linky}{Repository}} \ExternalLink
    \item \href{https://zenodo.org/records/10809833}{\color{linky}{Archive}} \ExternalLink
  \end{itemize}

  \vspace{2mm}

  \par\noindent\hrulefill\par

  \vspace{2mm}

  {\bfseries Editor:} \href{https://rmeli.github.io}{Rocco Meli} \ExternalLink\,\orcidlink{0000-0002-2845-3410} \\
  \vspace{1mm}
    {\bfseries Reviewers:}
  \begin{itemize}
  \setlength\itemsep{0em}
    \item \href{https://github.com/AlexBuccheri}{@AlexBuccheri}
    \item \href{https://github.com/katherineding}{@katherineding}
    \end{itemize}
    \vspace{2mm}

  {\bfseries Submitted:} 02 December 2023\\
  {\bfseries Published:} TBD

  \vspace{2mm}
  {\bfseries License}\\
  Authors of papers retain copyright and release the work under a Creative Commons Attribution 4.0 International License (\href{http://creativecommons.org/licenses/by/4.0/}{\color{linky}{CC BY 4.0}}).

  \end{flushleft}
}

\vspace{2.5em}

\hypertarget{summary}{%
\section{Summary}\label{summary}}

The Hofstadter model successfully describes the behavior of non-interacting quantum particles hopping on a lattice coupled to a gauge field, and hence is ubiquitous in many fields of research, including condensed matter, optical, and atomic physics. Motivated by this, we introduce HofstadterTools (\href{https://hofstadter.tools}{https://hofstadter.tools}), a Python package that can be used to analyze the energy spectrum of a generalized Hofstadter model, with any combination of hoppings on any regular Euclidean lattice. The package can be applied to compute key properties of the band structure, such as quantum geometry and topology, as well as plot Hofstadter butterflies and Wannier diagrams that are colored according to their Chern numbers.

\hypertarget{statement-of-need}{%
\section{Statement of need}\label{statement-of-need}}

\begin{figure}
	\includegraphics[width=\linewidth]{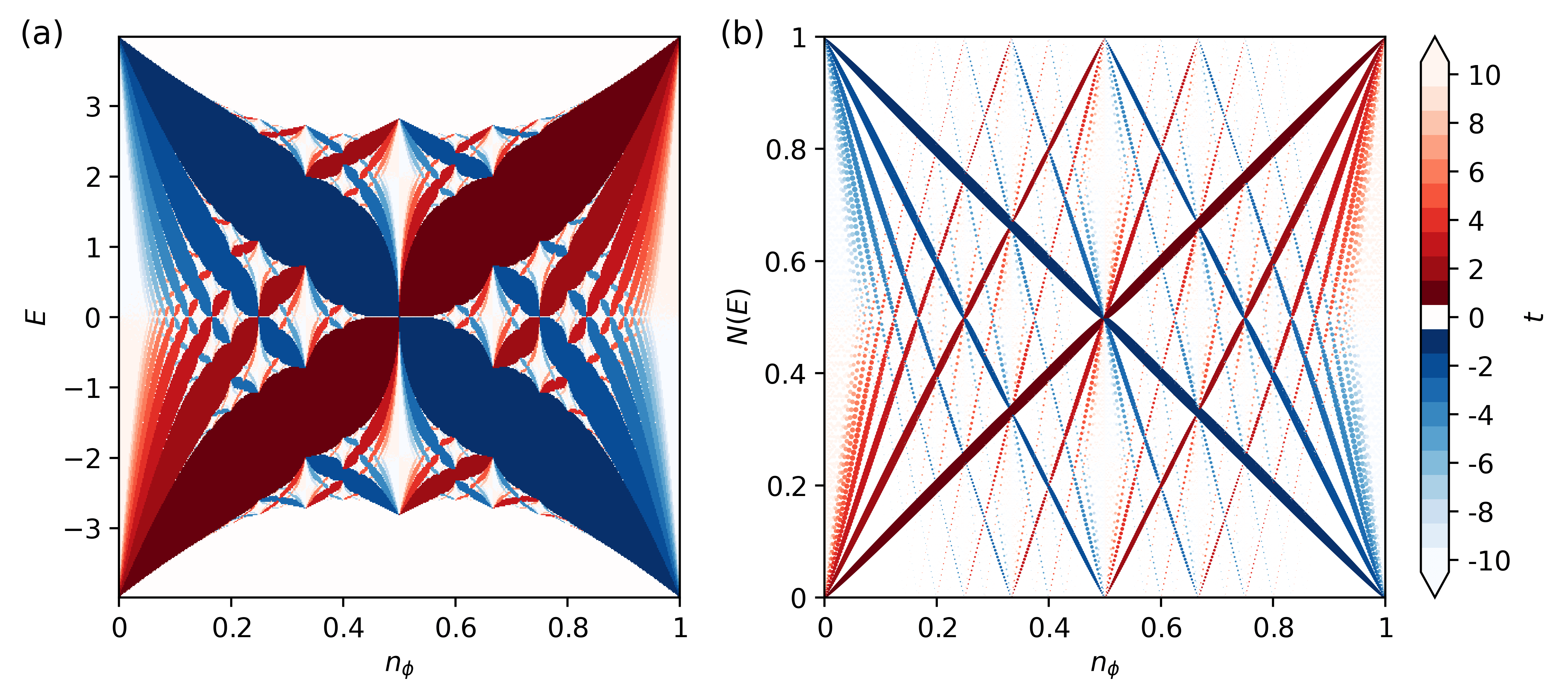}
	\caption{\label{fig:square}\textbf{Square Lattice.} (a)~Hofstadter butterfly and (b) Wannier diagram for the Hofstadter model defined with nearest-neighbor hoppings on the square lattice. (a)~The energy $E$, and (b)~the integrated density of states below the gap $N(E)$, are plotted as a function of flux density $n_\phi=BA_\mathrm{min}/\phi_0=p/499$, where $B$ is the perpendicular field strength, $A_\mathrm{min}$ is the area of a minimal hopping plaquette, $\phi_0$ is the flux quantum, and $p$ is an integer. The $r$-th gap is colored with respect to $t=\sum_{i=0}^r C_i$, where $C_i$ is the Chern number of band $i$. The size of the points in the Wannier diagram is proportional to the size of the gaps.~\cite{DiColandrea22}}
\end{figure}

\begin{figure}
	\includegraphics[width=\linewidth]{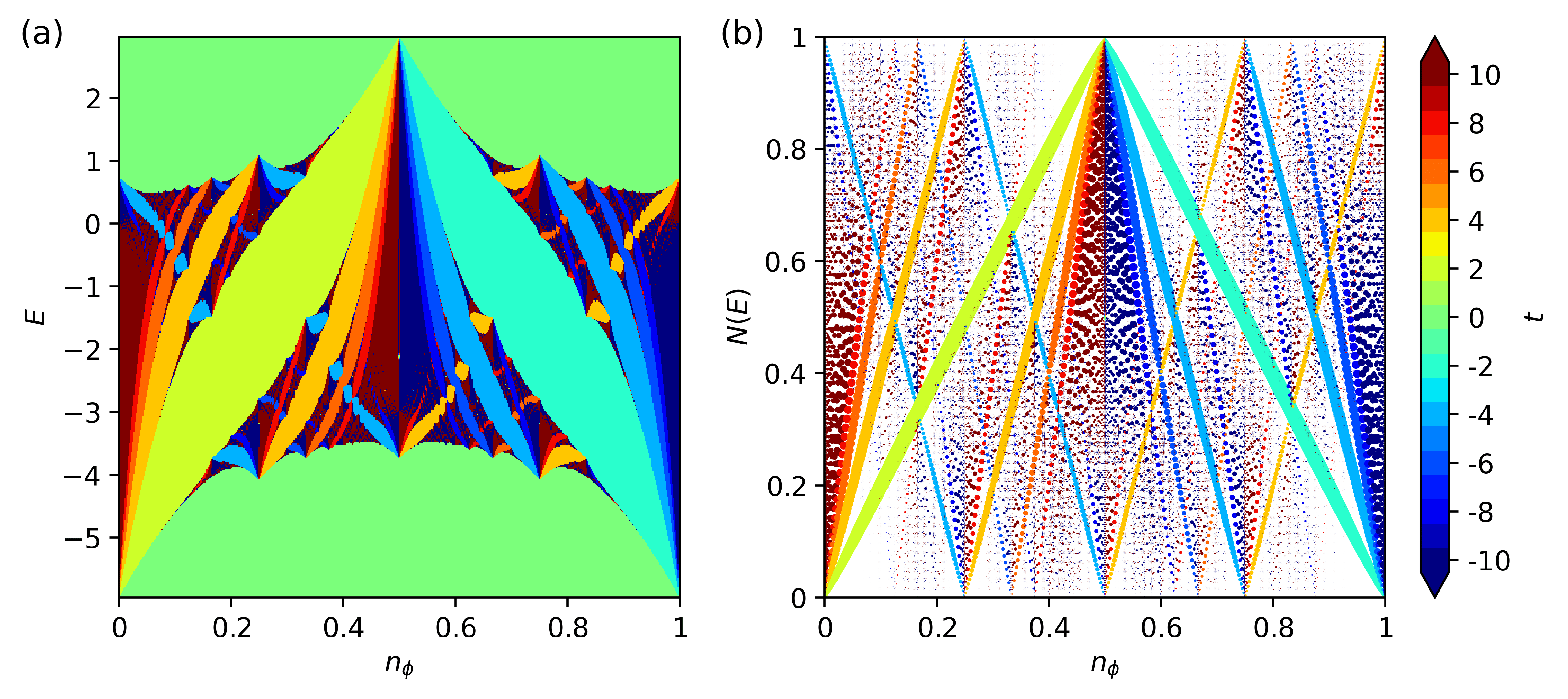}
	\caption{\label{fig:triangular}\textbf{Triangular Lattice.} (a)~Hofstadter butterfly and (b)~Wannier diagram for the Hofstadter model defined with nearest-neighbor hoppings on the triangular lattice.~\cite{Avron14}}
\end{figure}

\begin{figure}
	\includegraphics[width=\linewidth]{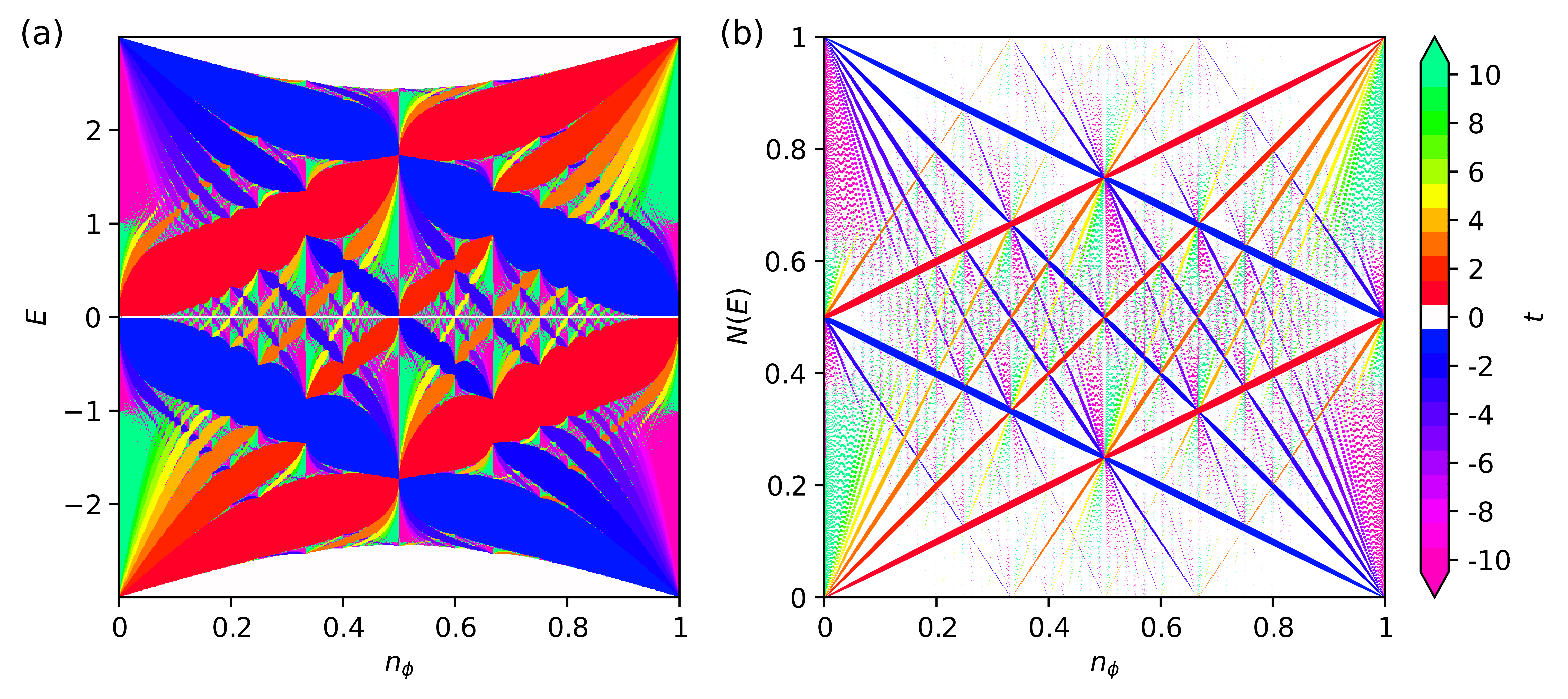}
	\caption{\label{fig:honeycomb}\textbf{Honeycomb Lattice.} (a)~Hofstadter butterfly and (b)~Wannier diagram for the Hofstadter model defined with nearest-neighbor hoppings on the honeycomb lattice.~\cite{Agazzi14}}
\end{figure}

\begin{figure}
	\includegraphics[width=\linewidth]{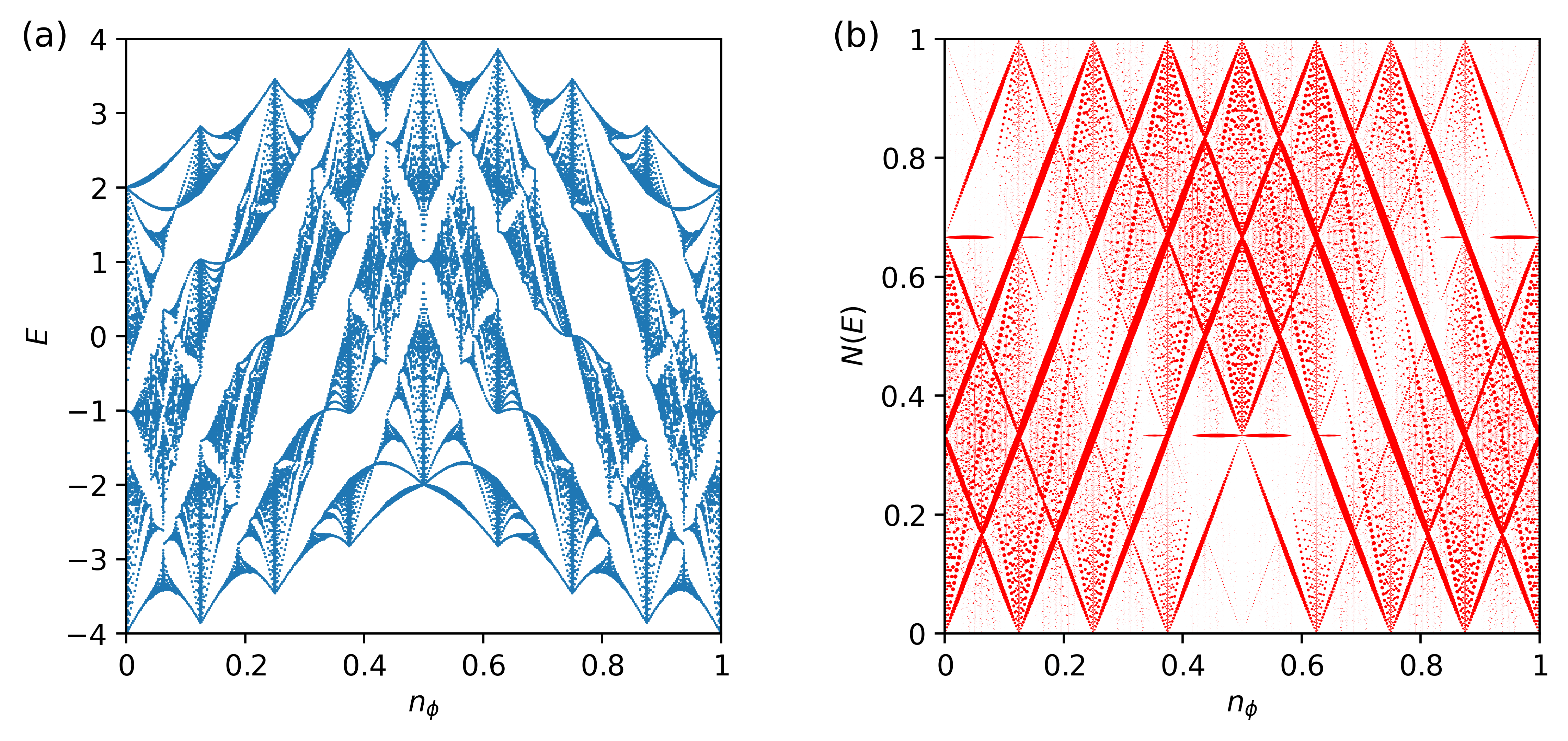}
	\caption{\label{fig:kagome}\textbf{Kagome Lattice.} (a)~Hofstadter butterfly and (b)~Wannier diagram for the Hofstadter model defined with nearest-neighbor hoppings on the kagome lattice.~\cite{Jing-Min09}}
\end{figure}

The purpose of HofstadterTools is to consolidate the fragmented theory and code relevant to the Hofstadter model into one well-documented Python package, which can be used easily by non-specialists as a benchmark or springboard for their own research projects. The Hofstadter model~\cite{Harper55, Azbel64, Hofstadter76} is an iconic tight-binding model in physics and famously yields a fractal energy spectrum as a function of flux density, as shown in Figs.~\ref{fig:square},~\ref{fig:triangular},~\ref{fig:honeycomb}, and~\ref{fig:kagome}. Consequently, it is often treated as an add-on to larger numerical packages, such as WannierTools~\cite{WannierTools}, pyqula~\cite{pyqula}, and DiagHam~\cite{DiagHam}, or simply included as supplementary code together with research articles~\cite{Bodesheim23}. However, the Hofstadter model's generalizability, interdisciplinary appeal, and recent experimental realization, motivates us to create a dedicated package that can provide a detailed analysis of its band structure, in the general case.

\begin{enumerate}
\item \textbf{Generalizability.} The Hofstadter model was originally studied in the context of electrons hopping in a periodic potential coupled to a perpendicular magnetic field. However, the model transcends this framework and can be extended in numerous directions. For example, the Peierls phases that arise in the Hamiltonian due to the magnetic field~\cite{Peierls33} can also be generated using artificial gauge fields~\cite{Goldman14} or Floquet modulation~\cite{Eckardt17}. Moreover, the full scope of the Hofstadter model is still being revealed, with papers on its application to hyperbolic lattices~\cite{Stegmaier22}, higher-dimensional crystals~\cite{DiColandrea22}, and synthesized materials~\cite{Bodesheim23}, all published within the last couple of years.

\item \textbf{Interdisciplinary appeal.} Owing to its generalizability, interest in the Hofstadter model goes beyond its well-known connection to condensed matter physics and the quantum Hall effect~\cite{Avron03}. In mathematics, for example, the difference relation arising in the solution of the Hofstadter model, known as the Harper equation~\cite{Harper55}, is a special case of an "almost Mathieu operator", which is one of the most studied types of ergodic Schrödinger operator~\cite{Simon00, Avila09}. Moreover, in other branches of physics, the Hofstadter model has growing relevance in a variety of subfields, including: cold atomic gases~\cite{Cooper19}, acoustic metamaterials~\cite{Ni19}, and photonics~\cite{Zilberberg18}.

\item \textbf{Recent experimental realization.} Although the Hofstadter model was introduced last century~\cite{Peierls33, Harper55}, it has only been experimentally realized within the last decade. Signatures of the Hofstadter spectrum were first observed in moiré materials~\cite{Dean13} and optical flux lattices~\cite{Aidelsburger13}, and they have since been reproduced in several other experimental platforms~\cite{Cooper19, Ni19, Zilberberg18, Roushan17}. Not only does this spur recent theoretical interest, but it also increases the likelihood of experimental groups entering the field, with the need for a self-contained code repository that can be quickly applied to benchmark data and related computations.
\end{enumerate}

A prominent use-case of HofstadterTools is to facilitate the study of a rich landscape of many-body problems. The Hofstadter model is an infinitely-configurable topological flat-band model and hence, is a popular choice among theorists studying strongly-correlated phenomena, such as the fractional quantum Hall effect~\cite{Andrews20, Andrews21} and superconductivity~\cite{Shaffer21, Sahay23}. Since there is a relationship between the quantum geometry and topology of single-particle band structures and the stability of exotic strongly-correlated states~\cite{Jackson15, Andrews23, Ledwith22, Lee17, Tian23, Wang21}, HofstadterTools may be used to guide theorists who are researching quantum many-body systems. More broadly, we hope that HofstadterTools will find many interdisciplinary applications, and we look forward to expanding the package in these directions, with help from the community.

\vfill
\vfill

\clearpage
\newpage

\vfill
\vfill
\vfill

\clearpage
\newpage

\hypertarget{acknowledgements}{%
\section{Acknowledgements}\label{acknowledgements}}

We thank Gunnar Möller, Titus Neupert, Rahul Roy, Alexey Soluyanov, Michael Zaletel, Johannes Mitscherling, Daniel Parker, Stefan Divic, and Mathi Raja, for useful discussions. This project was funded by the Swiss National Science Foundation under Grant No.~\href{https://data.snf.ch/grants/grant/203168}{P500PT\_203168}, and supported by the U.S.~Department of Energy, Office of Science, Basic Energy Sciences, under Early Career Award No.~DE-SC0022716.

\bibliographystyle{unsrtnat}
\bibliography{hof}

\end{document}